\documentclass[prl,twocolumn,nofootinbib,superscriptaddress]{revtex4-1} 

\usepackage{pdfpages}

\makeatletter
\AtBeginDocument{\let\LS@rot\@undefined}
\makeatother

\usepackage{amsmath}
\usepackage{bbm}
\usepackage{graphicx}
\usepackage{color}
\usepackage{amssymb}
\usepackage{units}
\usepackage[normalem]{ulem}
\usepackage[colorlinks=true,linkcolor=blue,citecolor=blue,urlcolor=blue]{hyperref}
\usepackage{siunitx}

\newcommand{\bea}{\begin{eqnarray}}
\newcommand{\eea}{\end{eqnarray}}
\newcommand{\be}{\begin{equation}}
\newcommand{\ee}{\end{equation}}
\newcommand{\benn}{\begin{equation*}}
\newcommand{\eenn}{\end{equation*}}

\newcommand{\kBT}{k_\text{B}T}

\definecolor{light-gray}{gray}{0.9}
\definecolor{gris}{gray}{0.5}
\definecolor{DarkGreen}{rgb}{0.,0.4,0.4}
\definecolor{amber}{rgb}{1,0.75,0}
\definecolor{brown}{rgb}{0.65, 0.16, 0.16}

\begin{document}

\title{Andreev-Coulomb Drag in Coupled Quantum Dots}
\author{S. Mojtaba Tabatabaei}
\affiliation{Department of Physics, Shahid Beheshti University, G. C. Evin, 1983963113 Tehran, Iran}
\author{David S\'anchez}
\affiliation{Institute for Cross-Disciplinary Physics and Complex Systems IFISC (UIB-CSIC), E-07122 Palma de Mallorca, Spain}
\author{Alfredo Levy Yeyati}
\affiliation{Departamento de F\'isica Te\'orica de la Materia Condensada, Condensed Matter Physics Center (IFIMAC), and Instituto Nicol\'as Cabrera, Universidad Aut\'onoma de Madrid, 28049 Madrid, Spain\looseness=-1}
\author{Rafael S\'anchez}
\affiliation{Departamento de F\'isica Te\'orica de la Materia Condensada, Condensed Matter Physics Center (IFIMAC), and Instituto Nicol\'as Cabrera, Universidad Aut\'onoma de Madrid, 28049 Madrid, Spain\looseness=-1}
\date{\today}

\begin{abstract}
The Coulomb drag effect has been observed as a tiny current induced by both electron-hole asymmetry and interactions in normal coupled quantum dot devices. In the present work we show that the effect can be boosted by replacing one of the normal 
electrodes by a superconducting one. Moreover, we show that at low temperatures and for sufficiently strong
coupling to the superconducting lead, the Coulomb drag is dominated by Andreev processes, is robust against details of the system parameters and can be controlled with a single gate voltage. This mechanism can be distinguished
from single-particle contributions by a sign inversion of the drag current.
\end{abstract}
\maketitle

{\it Introduction.---} 
The possibility to induce a current in an unbiased electronic circuit by proximity with a nearby driven system is a measurable manifestation of electron-electron correlations and broken symmetries. For this reason, the Coulomb drag effect has attracted theoretical and experimental attention for many decades~\cite{Narozny2016}.
In extended samples with translational invariance, momentum exchange across conductors is the relevant drag mechanism~\cite{rojo}. Remarkably, when the samples are superconducting (e.g., Coulomb coupled bilayers or Josephson junction arrays) Cooper pairs must be taken into account~\cite{oreg,Huang1995}, leading to drag supercurrents~\cite{duan}, charge solitons~\cite{shimada}, exciton formation~\cite{bercioux} and depinning~\cite{wilkinson} (see also Ref.~\cite{averin} for a discussion of drag currents in normal tunnel junction arrays).

In contrast, in mesoscopic systems such as coupled quantum dots or point contacts,
which lack translational invariance, Coulomb drag is driven by energy transfer. Moreover,
finite drag currents with normal leads require the presence of energy-dependent tunneling rates breaking electron-hole and inversion symmetries~\cite{Mortensen2001,Levchenko2008,Moldoveanu2009,Sanchez2010,Hussein2012,Hussein2015,Kaasberg2016,Sierra2019,Zhou2019,He2020}. Although this leads to a rather tiny and uncontrolled effect strongly dependent on the system parameters, drag currents have been observed in nanoelectronic devices~\cite{Shinkai2009,Laroche2011,Laroche2014,Bischoff2015,Keller2016}.
The mesoscopic Coulomb drag can also be viewed as arising from rectifying nonequilibrium charge fluctuations. This makes the connection with exciting issues in modern quantum thermodynamics: energy harvesting~\cite{hotspots,holger,roche}, heat transport~\cite{Koski2015,transistor,Bhandari2018}, and rectifiers~\cite{khrapai,fabian}.

\begin{figure}[b]
\includegraphics[width=\linewidth]{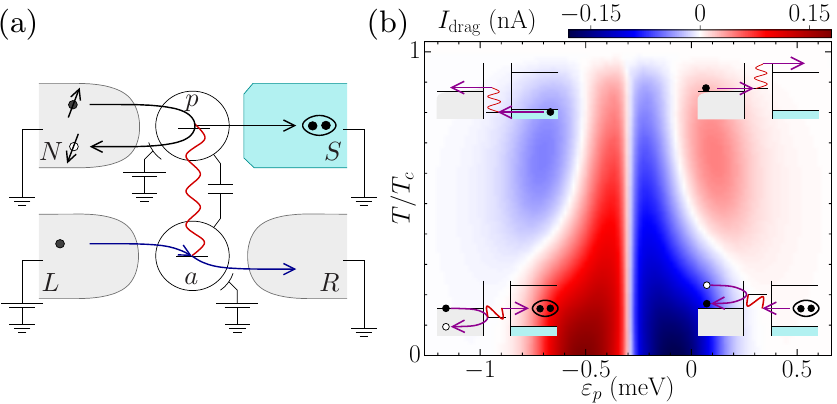}
\caption{\label{fig:scheme}Superconductivity induced Coulomb drag in hybrid dot structures. (a) Scheme of the capacitively coupled double quantum dot device analyzed in this work. A drag current is generated in the passive dot ($p$) connected to unbiased normal ($N$) and superconducting ($S$) electrodes due to the Coulomb interaction with electrons tunneling through the active dot, $a$, attached to voltage biased terminals ($L$ and $R$). In this diagram, an electron in $N$ is transformed into a Cooper pair in $S$ with a retroreflected hole. This Andreev process is correlated through the interaction (red line) with a charge fluctuation in the active subsystem (blue arrow). (b) Drag current $I_{\rm drag}$ as a function of the passive dot level $\varepsilon_p$ and temperature $T$ (normalized with the critical temperature $T_c$). Cooper pairs and quasiparticles contribute with opposite signs.
At low temperature, the current is given by Andreev processes (lower insets). By increasing temperature, a crossover occurs where they coexist with quasiparticle tunneling (upper insets). Parameters (in meV): $\Delta(T{=}0)=0.2$, $\Gamma_S=\Gamma_N=\Gamma_L=\Gamma_R=0.05$, $V_{\rm{bias}} =5$, $\varepsilon_a{=}0$, and $U_p=5U_{ap}=0.5$.} 
\end{figure}

In this Letter, we explore how the mesoscopic Coulomb drag is affected by superconducting correlations. Let us consider a proximitized nanostructure~\cite{hybrid-review} like the one depicted in Fig.~\ref{fig:scheme}(a).  It consists of two capacitively coupled quantum dots~\cite{Chan2002,Hubel2007,Mcclure2007}: the active dot is coupled to two voltage-biased normal reservoirs, whereas the passive one is coupled to a normal and a superconducting lead (hence breaking inversion symmetry).
Due to the presence of a superconducting gap, the tunneling rates exhibit a well defined energy dependence which has proven useful for quasiparticle turnstiles~\cite{vanZanten2016} and refrigeration~\cite{nahum,leivo,refrig,hsu}. 

Superconductivity gives rise, however, to an additional and totally different drag mechanism via the Andreev reflection of an electron into a hole at the passive quantum dot~\cite{alvaro}.
Mediated by Coulomb interactions, charge fluctuations in the active dot correlates with the pairing processes, hence breaking electron-hole symmetry, as schematically depicted in the lower insets of Fig.~\ref{fig:scheme}(b). A key role is played by the coherent superposition of the passive dot charge states induced by pairing and which can be controlled by external gate voltages. Tuning the position of the passive dot level, the drag current would display a transition from electron- to hole-like dominated behavior. As we show in this work, the Andreev drag mechanism dominates at low temperature $T$ for realistic system parameters, and could be distinguished from the conventional (single particle) processes by a drag current sign reversal in state-of-the-art experiments. 

{\it Model and methods.---} We model the device depicted in Fig.~\ref{fig:scheme}(a) with a Hamiltonian of the form $\hat{H}=\hat{H}_{\rm dqd} + \hat{H}_{\rm leads} + \hat{H}_t$, where
\begin{equation}
    \hat{H}_{\rm dqd}=\underset{\alpha}{\sum}\varepsilon_{\alpha}\hat{n}_{\alpha}+U_{p}\hat{n}_{p,\uparrow}\hat{n}_{p,\downarrow}+U_{ap}\hat{n}_{a}\hat{n}_{p},
\end{equation}
is the double quantum dot Hamiltonian with active ($\alpha=a$) and passive ($\alpha=p$) dots, energy levels $\varepsilon_{\alpha}$, intra- ($U_{p}$) and interdot ($U_{ap}$) charging energies and number operators $\hat{n}_{p}=\Sigma_{\sigma}\hat{n}_{p,\sigma}=\Sigma_{\sigma}\hat{d}_{p,\sigma}^{\dagger}\hat{d}_{p,\sigma}$ 
with spin $\sigma=\{\uparrow,\downarrow\}$ and
$\hat{n}_a=\hat{d}^{\dagger}_a\hat{d}_a$, where $\hat{d}_{p,\sigma}$ and $\hat{d}_a$ are the electron annihilation operators in passive and active dots (note that, for simplicity, the active dot is described in terms of spinless electrons as this degree of freedom does not play a role in the drag physics).
$\hat{H}_{\rm leads}$
and $\hat{H}_t$ are the uncoupled leads and the lead-dot tunneling Hamiltonians, respectively.
While the passive dot is coupled to a normal ($N$) and a superconducting ($S$) lead, the active dot is connected to two normal ones ($L$ and $R$).
The corresponding tunneling rates are denoted by $\Gamma_\beta$,
where $\beta=N,S,L,R$ labels the terminals with 
chemical potential $\mu_\beta$.
Since the passive dot is in equilibrium we set $\mu_N=\mu_S=0$, while the active dot is voltage biased as $\mu_L=-\mu_R=eV_{\rm{bias}}/2$.

In order to obtain the transport properties for this model we apply two complementary approaches. For the numerical calculations, we use a nonequilibrium Green's functions (NEGF) formalism including up to second-order diagrams for the interaction self-energies and taking into account the BCS temperature dependence for the $S$ lead gap. Additionally, we use a master equation description to analytically identify the relevant processes responsible for the drag current. This latter approach is valid in two limiting cases: (i) the drag is carried by Cooper pairs (assuming the infinite limit for the superconducting gap $\Delta$), and (ii) pairing is neglected and drag is due to single-quasiparticle processes only \cite{SM}.

The drag current as a function of temperature and passive dot level obtained using NEGF is shown in Fig.~\ref{fig:scheme}(b).
The chosen parameters
correspond to realistic values that can be achieved in current experimental setups \cite{hybrid-review}.
As a first remark, we notice that the drag current can reach values of the order of $\unit[0.1]{nA}$, which can be easily detected experimentally~\cite{Keller2016}. Secondly, we observe four different regions depending on the sign of the drag current. At low temperatures, this sign changes when $\epsilon_p$ crosses the electron-hole symmetric point $\varepsilon_p^{\rm eh}=-(U_p+U_{ap})/2$, turning from positive to negative as the passive dot is depopulated. At higher temperatures an additional sign change occurs, while the size of the drag current is significantly reduced, and eventually disappears when the gap is closed for $T>T_c$. As discussed below, this change of behaviour is physically connected to the microscopic mechanisms yielding the Coulomb drag, pinpointing a crossover from the Andreev drag regime to the quasiparticle tunneling regime as temperature is raised. 

\begin{figure}[b]
\includegraphics[width=\linewidth]{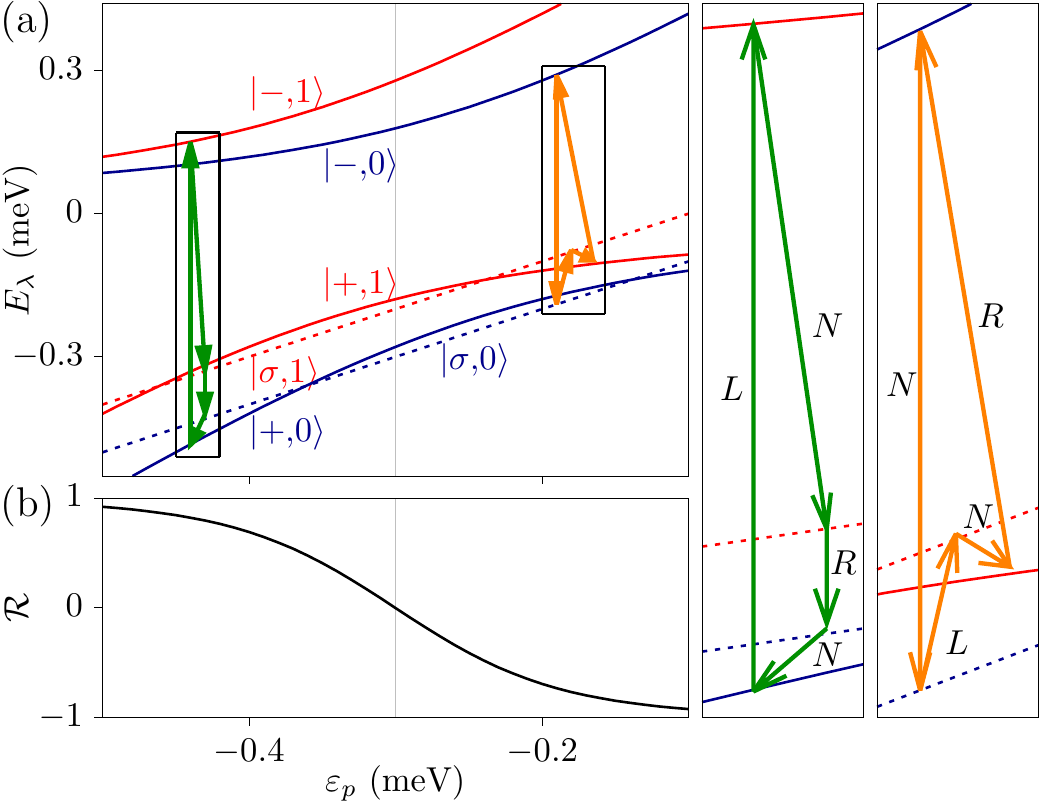}
\caption{\label{fig:pairing}Andreev-Coulomb drag mechanism. (a) Pairing induces avoided crossings in the even states $|\pm,n\rangle$.
Two cyclic processes are highlighted which correspond to  Eq.~\eqref{eq:cycle}, with $n=1$, $n'=0$ (orange) and $n=0$, $n'=1$ (green). These cycles are assisted by charge fluctuations in the active dot and lead to finite drag currents. 
(b) The imbalance of tunneled electrons and holes dictates the sign of the current by means of the asymmetry of the transition rates, parametrized here by ${\cal R}=(r_{+0}r_{-1}{-}1)/(r_{+0}r_{-1}{+}1)$, depending on the level position $\varepsilon_{p}$. The grey line in (a) and (b) marks the electron-hole symmetry point, $\varepsilon_p^{\rm eh}$.
Parameters as in Fig.~\ref{fig:scheme}(b).
}
\end{figure}

{\it Drag out of correlated Andreev reflection.---} We can gain insight on the low temperature behavior using the master equation approach with the transition rates calculated in the limit $\Delta,eV_{\rm{bias}}\gg\kBT$. In this approximate analysis, the relevant basis states are $|i,n\rangle$, where $i=\{0,\sigma,2\}$ denotes the occupation of the passive dot and $n=0,1$ is the number of electrons in the active one.
Pairing induced by coupling to the superconductor hybridizes the even-$i$ states in the passive dot, leading to ${|}{\pm}{,}n\rangle={\cal N}_{\pm,n}^{-1}(A_{\pm,n}{|}0{,}n\rangle-\Gamma_S{|}2{,}n\rangle)$, with
$A_{\pm,n}=\tilde\varepsilon_n\pm\sqrt{\tilde\varepsilon_n^2{+}\Gamma_S^2}$, $\tilde\varepsilon_n=\varepsilon_p+nU_{ap}+U_p/2$,
and the normalization factor ${\cal N}_{\pm,n}$.
The odd states ${|}{\sigma}{,}n\rangle$ remain uncoupled.
Importantly, the even superpositions and the eigenenergies $E_{\pm,n}=n\varepsilon_a+A_{\mp,n}$,
$E_{\sigma n}=\varepsilon_{p}+n(\varepsilon_{a}+U_{ap})$ depend on the occupation of the active dot, see Fig.~\ref{fig:pairing}(a).

In the weak tunneling regime, the transition rates between the different states $|\kappa\rangle{\rightarrow}|\lambda\rangle$ are given by $\Gamma_{\lambda\kappa}^{\alpha\beta}={\cal G}_{\lambda\kappa}^{\alpha\beta}f_\beta(E_\lambda{-}E_\kappa)$
when involving an electron tunneling from terminal $\beta$ into quantum dot $\alpha$, and $\gamma_{\lambda\kappa}^{\alpha\beta}={\cal J}_{\lambda\kappa}^{\alpha\beta}[1-f_\beta(E_\kappa{-}E_\lambda)]$
when involving a hole~\cite{Sauret2004,Eldridge2010,cpscooling}. Here, ${\cal G}_{\lambda\kappa}^{\alpha\beta}=\Gamma_\beta|\langle\lambda|\hat{\delta}_{\alpha}^\dagger|\kappa\rangle|^2$ are the leading-order tunneling rates with $\hat{\delta}_{a} \equiv \hat{d}_a$, $\hat{\delta}_p \equiv \sum_{\sigma} \hat{d}_{p,\sigma}$ and $f_\beta(E)=\{1+\exp[(E{-}\mu_\beta)/\kBT]\}^{-1}$ the Fermi-Dirac function. For ${\cal J}_{\lambda\kappa}^{\alpha\beta}$, one replaces $\hat{\delta}_{\alpha}^\dagger$ by $\hat{\delta}_{\alpha}$ in the expression for ${\cal G}_{\lambda\kappa}^{\alpha\beta}$. 

At low temperatures, transitions $|\kappa\rangle{\rightarrow}|\lambda\rangle$ involving terminal $N$ are energetically suppressed if $E_\lambda-E_\kappa\gg\mu_N=0$. Noticing that $E_{\pm n}-E_{\sigma n}=U_p/2\mp\sqrt{\tilde\varepsilon_n^2{+}\Gamma_S^2}$, we see that transitions to $|{-}{,}n\rangle$ are only possible if mediated by a fluctuation in the active dot. Due to the interaction $U_p$, the energies of the $|{+}{,}n\rangle$ and $|{\sigma}{,}n\rangle$ states might cross, as shown in Fig.~\ref{fig:pairing}(a). 
Hence, for a constant $n$ the passive dot is in a statistical mixture of states ${|}{+}{,}n\rangle$ and ${|}{\sigma}{,}n\rangle$. Then, electrons tunneling from $L$ to $R$ change the charge $n$ and consequently the passive dot undergoes transitions between different states. In particular, depending on which is the ground state, the transitions ${|}{+}{,}n\rangle\leftrightarrow{|}{-}{,}n{\pm}1\rangle$ and ${|}{\sigma}{,}n\rangle\leftrightarrow{|}{\sigma}{,}n{\pm}1\rangle$ are important for unblocking the dynamics. Furthermore, they enable cycles of the form 
\begin{equation}
\label{eq:cycle}
|\sigma{,}n\rangle{\rightarrow}|{+}{,}n\rangle{\leftrightarrow}|{-}{,}n'\rangle{\rightarrow}|{\sigma}{,}n'\rangle{\leftrightarrow}|{\sigma}{,}n\rangle,
\end{equation}
as sketched in Fig.~\ref{fig:pairing}(a) by
the green (for $n'=n+1$) and orange (for $n'=n-1$) arrows.
They include two transitions in terminal $N$ involving different superpositions at different occupations of the active dot, $|{+}{,}n\rangle$ and $|{-}{,}n'\rangle$. Thus, the weights in these superpositions are changed in the processes due to both pairing and Coulomb interactions. These are the minimal cycles that allow for a net charge exchange between $N$ and $S$ and ultimately lead to a drag current. 

The drag effect requires the electron-hole symmetry to be dynamically broken~\cite{Sanchez2010}. In our system, transitions ${|}\sigma{,}n\rangle{\rightarrow}{|}{+}{,}n\rangle$ and ${|}{-}{,}n\rangle{\rightarrow}{|}\sigma{,}n\rangle$ can both be due to the tunneling of either an electron or a hole into the passive dot. Whether electron or hole transport dominates depends in each case on the relative weight of the even states via the ratios $r_{+n}{=}{\cal G}_{+n,\sigma n}^{pN}/{\cal J}_{+n,\sigma n}^{pN}{=}\Gamma_S^2/A_{+n}^2$ and $r_{-n}{=}{\cal G}_{\sigma n,-n}^{pN}/{\cal J}_{\sigma n,-n}^{pN}{=}A_{-n}^2/\Gamma_S^2$.
In the cycles of Eq.~\eqref{eq:cycle}, more electrons than holes will on average be transferred if $r_{+n}r_{-n'}>1$. Remarkably, these ratios can be controlled by tuning the passive dot energy level with a gate voltage. For $r_{\pm n} > 1$ electron tunneling will occur at a higher rate than hole tunneling if $\tilde\varepsilon_n<0$. In this way if $\tilde\varepsilon_0{<}\tilde\varepsilon_1{<}0$ (like the green example in Fig.~\ref{fig:pairing}), we have ${A}_{+,n}^2{<}\Gamma_S^2{<}{A}_{-,n}^2$. Then most likely, two electrons tunnel from $N$ into the passive dot along with a Cooper pair created in the superconductor, giving rise to a positive drag current. In the opposite case $\tilde\varepsilon_0>0$ (like for the orange cycle in Fig.~\ref{fig:pairing}), the two transitions are hole-like and the drag current is accordingly reversed. As shown in Fig.~\ref{fig:pairing}(b), $r_{+n}r_{-n'}-1$ is monotonic when tuning $\varepsilon_p$ across the intermediate region. It changes sign at $\varepsilon_p^{\rm eh}$, together with $I_{\rm drag}$ [see Fig.~\ref{fig:scheme}(b)]. At this point electron and hole processes occur with the same rate during the cycle and cancel each other, resulting in no net drag current. 

We stress that the pairing term is necessary for the drag current in the Andreev regime since (i) it splits the degeneracy points of even and odd states, and (ii) it breaks electron-hole symmetry by making $r_{\pm n}\neq1$. However, increasing the $S$ coupling has a counter-productive effect: in the limit $\Gamma_S\gg\tilde\varepsilon_n$ we find $r_{\pm n}\rightarrow1$, i.e., tunneling becomes electron-hole symmetric and the drag effect accordingly vanishes.

Note that this drag effect differs from the normal one~\cite{Sanchez2010,Kaasberg2016,Sierra2019}
in that both the existence of the drag current
and its sign are determined by $\varepsilon_p$,
while in normal systems both depend
on energy-resolved tunneling asymmetries.
Our device can also host such conventional drag effect
when current is carried by quasiparticles,
as discussed below. Strikingly enough,
the sign of their contribution is also well defined in terms of $\varepsilon_p$ but is opposite to that of the Andreev-Coulomb drag processes. This would be useful for an experimental distinction of the two mechanisms.

\begin{figure}[t]
\includegraphics[width=\linewidth]{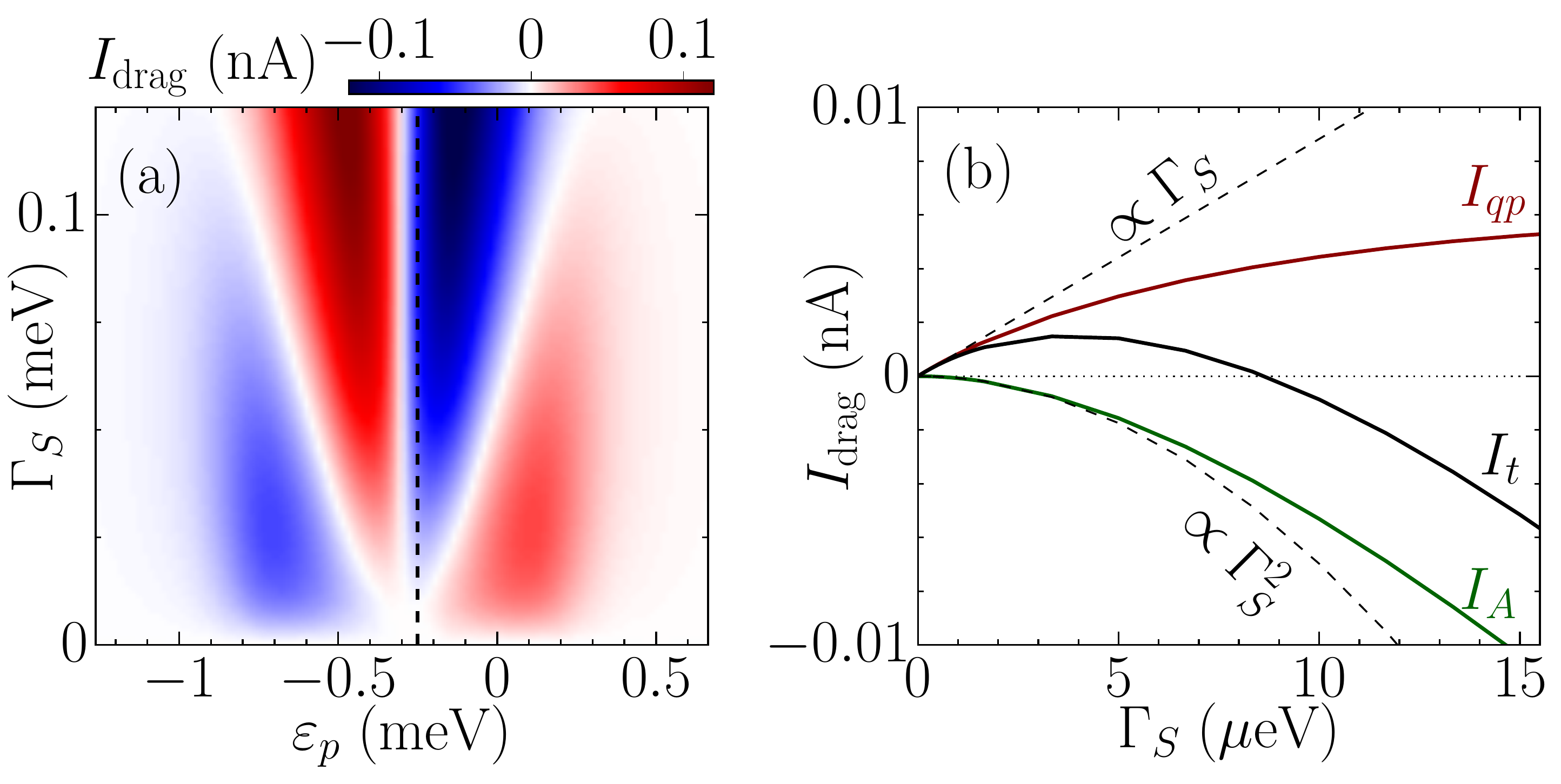}
\caption{\label{fig:optim}Unraveling the Andreev and quasiparticle contributions to the drag current. (a) Drag current as a function of $\Gamma_S$ and $\varepsilon_p$. (b) Different contributions from the single quasiparticle ($I_{\rm qp}$) and pair tunneling ($I_{\rm A}$) currents to the total drag current ($I_{\rm t}$). Dashed lines in (b) are fitting curves to the $\Gamma_S$ and $\Gamma_S^2$ lines. Parameters as in Fig.~\ref{fig:scheme}, except for $T=2T_c/3$. 
}
\end{figure}

{\it Quasiparticle drag.---} Let us now consider a regime for which pair tunneling can be neglected (e.g., high temperatures and quantum dot levels close to the gap edges)
and therefore the number of electrons in the passive dot is well defined. In a qualitative analysis, we can restrict the charge of the passive dot to fluctuate between 0 or 1. In the weak coupling regime, transport through the $S$ barrier consists of single-electron tunneling events whose rate $\Gamma_{Sn}^{\rm qp}\propto \nu_S(\varepsilon_p+nU_{ap})$ has an explicit energy dependence given by the superconductor density of states $\nu_S(E)$~\cite{SM}.
Since the rate is sensitive to the charge state of the active dot $n$ due to the Coulomb interaction $U_{ap}$, this situation is then analogous to that of Ref.~\cite{Sanchez2010}, where the drag effect appears in the presence of energy-dependent barriers. 
Here, such dependence is provided by the superconducting gap, even if all other couplings $\Gamma_\beta$ are constant. Thus, a drag current appears proportional to $\Gamma_{S1}^{\rm qp}-\Gamma_{S0}^{\rm qp}$, which depends on $\varepsilon_p$. 
For instance, if $-\Delta<\varepsilon_p<\Delta$ and $\varepsilon_p+U_{ap}>\Delta$, we have $\Gamma_{S1}^{\rm qp}\gg\Gamma_{S0}^{\rm qp}$, and consequently electrons tunnel from $N$ when $n=0$ and into $S$ over the gap once the active dot gets occupied (thereby $n$ changing to $1$). This causes a positive drag current. For $\varepsilon_p<-\Delta$ and $-\Delta<\varepsilon_p+U_{ap}<\Delta$, the same happens for holes tunneling below the gap ($n$ changing from 1 to 0) and the current becomes negative. 
Both cases are illustrated with the upper insets in Fig.~\ref{fig:scheme}(b). As in the case of the pairing contribution, the current changes sign at $\varepsilon_p=\varepsilon^{\rm eh}_p$.

{\it Andreev vs.\ quasiparticle drag.---}
The two limiting cases discussed above lead to opposite contributions to the drag current for the same configuration of the dot levels. Their competition will then determine the overall sign of the generated current, for instance, as a function of temperature as already shown in Fig.~\ref{fig:scheme}(b).
The value of $\varepsilon_p$ relative to the crossover depends on temperature since the BCS gap changes with temperature.
We furthermore investigate this competition as a function of the pairing parameter $\Gamma_S$ in Fig.~\ref{fig:optim}. Since the leading contribution for the quasiparticle tunneling is sequential, it is of the order $\Gamma_S \Gamma_N$. However, the Andreev contribution relies on higher order processes and depends on $\Gamma_S$ (through $r_{\pm n}$) as $(\Gamma_S\Gamma_N)^2$. Hence, for small values of $\Gamma_S$ the quasiparticle contribution dominates. As the pairing term increases, the Andreev-Coulomb mechanism starts to govern the drag effect.
This is reflected in a sign change of the drag current as a function of $\Gamma_S$ for a fixed dot level, see Fig.~\ref{fig:optim}(a).

This interpretation is confirmed by means of an energy-resolved separation of the different contributions to the drag current within the NEGF calculations. Let $I_{\rm A}$ be the subgap term, which we expect to be mainly due to  Andreev processes, and $I_{\rm qp}$ the contribution over and below the gap, mainly due to quasiparticles (see Ref.~\cite{SM} for further details). These are plotted in Fig.~\ref{fig:optim}(b) for $\varepsilon_p>\varepsilon_p^{\rm eh}$ [along the dashed line in Fig.~\ref{fig:optim}(a)]. It indeed shows that for small $\Gamma_S$, $I_{\rm A}$ is quadratic and negative, while $I_{\rm qp}$ is linear and positive, as expected for the Andreev and quasiparticle processes discussed above.

\begin{figure}[t]
\includegraphics[width=\linewidth]{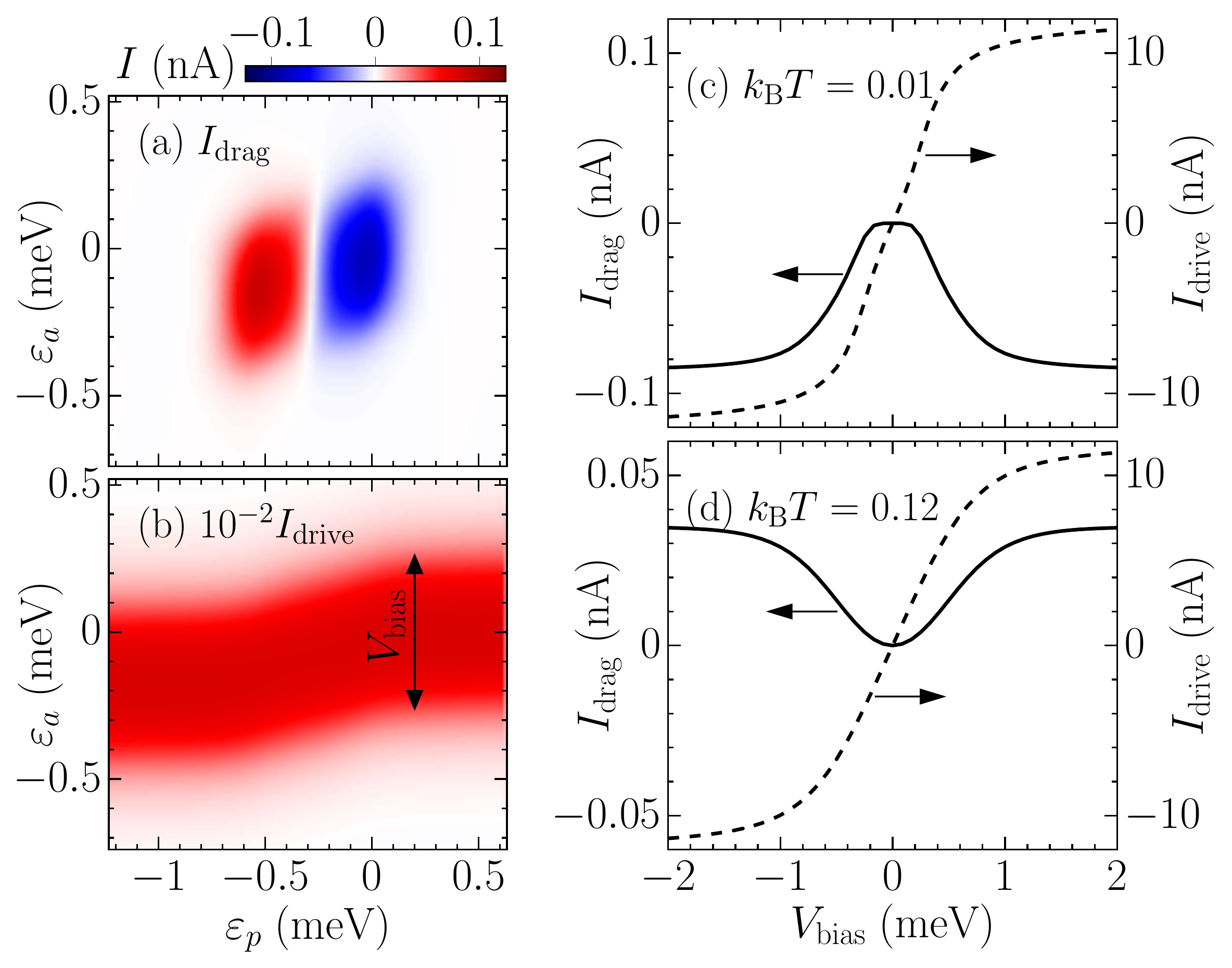}
\caption{\label{fig:Vbias}Drag and drive currents as functions of experimentally tunable gate and bias voltages. (a), (b) Mutual backaction of the drag and drive systems as $\varepsilon_a$ and $\varepsilon_p$ scan the degeneracy points of the stability diagram for $k_{\rm B}T=\unit[0.01]{meV}$ and $V_{\rm bias}=\unit[0.6]{meV}$. (c), (d) The drag current does not change sign at zero bias voltage, proving its nonlinearity, here for $\varepsilon_p=0$. (c) At low temperatures, the Andreev-Coulomb drag current is blocked at low voltages. Lifting of the blockade affects the drive response. (d) At higher temperatures, these features are washed out. Quasiparticle contributions change the sign of the current. Other parameters are as in Fig.~\ref{fig:scheme} and $k_B T_c = 0.15$~meV.}
\end{figure}

{\it Practical considerations.---} In a typical experiment, one could tune the gate voltages that control $\varepsilon_a$ and $\varepsilon_p$, as well as the bias voltage $V_{\rm bias}$ applied to the active dot. The drag effect manifests as a correlation between the currents through the active and passive dots, see Figs.~\ref{fig:Vbias}(a) and (b). These correlations induced by $U_{ap}$ are enhanced in the vicinity of the charge stability diagram degeneracy points, which give rise to an upward shift in the drive current, cf. Fig.~\ref{fig:Vbias}(b). This behaviour is typical of fluctuation-induced transport in normal double dot systems and leads to a current spot in the passive system whose sign depends on the potential in the passive system~\cite{Bischoff2015,holger,Keller2016}. Remarkably, for the Andreev-Coulomb drag effect discussed here the induced current feature is split into electron-like and hole-like contributions with opposite signs, as shown in Fig.~\ref{fig:Vbias}(a).

The dependence on the bias voltage plotted in Figs.~\ref{fig:Vbias}(c) and (d) shows that the drag current appears at the onset of nonlinearities in the drive current for large enough voltages, as expected for active systems with no energy-dependent tunneling rates~\cite{Sanchez2010,Sierra2019}. This is particularly clear at low temperatures where $I_{\rm drag}$ is blocked for low voltages, see Fig.~\ref{fig:Vbias}(c). The size of the blockade region scales linearly with the Coulomb interaction, $U_{\rm ap}$. At higher temperatures [Fig.~\ref{fig:Vbias}(d)] the blockade is smeared out. Further, we observe that the drag current changes sign with temperature due to the dominance of either pair-correlated [Fig.~\ref{fig:Vbias}(c)] or single-electron [Fig.~\ref{fig:Vbias}(d)] procesess.
This  behaviour  is  absent  in  coupled superconducting wires or layers when their drag effect is due to supercurrents~\cite{duan}. In Josephson junction arrays, the drag current can change its sign when the superconducting gap is suppressed~\cite{shimada,wilkinson}. However, in these devices the induced current flows in the same (opposite) direction as the drive current when transport is due to pairs (quasiparticles), while in our case it is independent of the direction of the drive and can be controlled by a local gate electrode.

{\it Conclusions.---}
We have shown that pair tunneling in interacting quantum dots coupled to normal and superconducting electrodes leads to a novel drag mechanism. This Andreev-Coulomb drag can be distinguished
from the conventional single-electron mechanism by means of a sign change in the drag current. Using two different theoretical methods, we have evaluated the drag currents showing that the Andreev mechanism is a robust effect and dominates at sufficiently small temperatures and strong coupling to the superconducting electrode. Our estimations, based on realistic parameters, indicate that the effect would be significantly strong to be detected using nowadays experimental techniques.

\acknowledgements
\textit{Acknowledgements.---} Work funded by MICINN (Ministerio de Ciencia e Innovaci\'on) under Grants Nos.\ MAT2017-82639, RYC2016-20778, FIS2017-84860-R and PID2019-110125GB-I00, and the “María de Maeztu” Programme for Units of Excellence in R\&D (Grant Nos.\ MDM2017-0711 and CEX2018-000805-M).

\bibliographystyle{apsrev}

\pagebreak
\widetext
\clearpage


\section*{Supplementary material}

# Supplementary materials for "Andreev-Coulomb Drag in Coupled Quantum Dots"

S. Mojtaba Tabatabaei,[1] David Sánchez,[2] Alfredo Levy Yeyati,[3] and Rafael Sánchez[3]

[1]*Department of Physics, Shahid Beheshti University, G. C. Evin, 1983963113 Tehran, Iran*
[2]*Institute for Cross-Disciplinary Physics and Complex Systems IFISC (UIB-CSIC), E-07122 Palma de Mallorca, Spain*
[3]*Departamento de Física Teórica de la Materia Condensada, Condensed Matter Physics Center (IFIMAC), and Instituto Nicolás Cabrera, Universidad Autónoma de Madrid, 28049 Madrid, Spain*

## S-I. THE MODEL

The total Hamiltonian is $\hat{H} = \hat{H}_\text{dqd} + \hat{H}_\text{leads} + \hat{H}_t$. The first term is the Hamiltonian for the double quantum dot [Eq. (1) in the main text]:

$$\hat{H}_\text{dqd} = \sum_\alpha \varepsilon_\alpha \hat{n}_\alpha + U_p \hat{n}_{p,\uparrow} \hat{n}_{p,\downarrow} + U_{ap} \hat{n}_a \hat{n}_p. \tag{S1}$$

Here, the number operators are $\hat{n}_p = \Sigma_\sigma \hat{n}_{p,\sigma} = \Sigma_\sigma \hat{d}^\dagger_{p,\sigma} \hat{d}_{p,\sigma}$ with spin $\sigma = \{\uparrow, \downarrow\}$ and $\hat{n}_a = \hat{d}^\dagger_a \hat{d}_a$ where $\hat{d}_{p,\sigma}$ and $\hat{d}_a$ denote the electron annihilation operators in passive ($\alpha = p$) and active ($\alpha = a$) dots with energy $\varepsilon_\alpha$ (we take spinless electrons for $\alpha = a$ since the spin degree of freedom in the active subsystem does not change the main physics of the problem). The Hamiltonian of the electrodes is given by

$$\hat{H}_\text{leads} = \sum_k \varepsilon_{k,L} \hat{c}^\dagger_{k,L} \hat{c}_{k,L} + \sum_k \varepsilon_{k,R} \hat{c}^\dagger_{k,R} \hat{c}_{k,R} + \sum_{k,\sigma} \varepsilon_{k,N} \hat{c}^\dagger_{k,N,\sigma} \hat{c}_{k,N,\sigma} + \sum_{k,\sigma} \varepsilon_{k,S} \hat{c}^\dagger_{k,S,\sigma} \hat{c}_{k,S,\sigma} + \sum_k \Delta \left( \hat{c}^\dagger_{k,S,\uparrow} \hat{c}^\dagger_{k,S,\downarrow} + \text{h.c.} \right), \tag{S2}$$

where $\hat{c}_k$ is the annihilation operator for electrons with energy $\varepsilon_k$ and $\Delta$ is the order parameter in the superconducting lead. The passive dot is coupled to normal ($N$) and superconductor ($S$) electrodes. In the drag configuration, these two electrodes have the same chemical potential $\mu$. Without loss of generality, we take the energy reference at $\mu_N = \mu_S = 0$. Furthermore, the active dot is attached to two normal electrodes ($L$ and $R$) through which a symmetric bias voltage is applied as $\mu_L = -\mu_R = eV_\text{bias}/2$, where $e$ is the electron charge. As a consequence, the tunneling Hamiltonian becomes

$$\hat{H}_t = \sum_k \left( t_L \hat{d}^\dagger_a \hat{c}_{k,L} + t_R \hat{d}^\dagger_a \hat{c}_{k,R} + \text{h.c.} \right) + \sum_{k,\sigma} \left( t_N \hat{d}^\dagger_{p,\sigma} \hat{c}_{k,N,\sigma} + t_S \hat{d}^\dagger_{p,\sigma} \hat{c}_{k,S,\sigma} + \text{h.c.} \right), \tag{S3}$$

where $t_\beta$ for $\beta = L, R, N, S$ are the dot-lead tunnel couplings. In the following, we consider the wide-band approximation, in which case the tunnel hybridization strength is given by $\Gamma_\beta = 2\pi |t_\beta|^2 \rho_0^\beta$, where $\rho_0^\beta$ is the corresponding electrode's density of states in its normal state.

## S-II. NONEQUILIBRIUM GREEN'S FUNCTIONS METHOD

Here, we give the details of calculating the drag current using the nonequilibrium Green's functions (NEGF) formalism [S1]. We consider the Hamiltonian of noninteracting double quantum dots, $\hat{H}_\text{dqd}(U_p = U_{ap} = 0)$, and the Hamiltonian of electrodes, $\hat{H}_\text{leads}$, as the unperturbed Hamiltonian and proceed by considering $\hat{H}_t$ and $\hat{H}_\text{int} = U_{ap} \hat{n}_a \hat{n}_p + U_p \hat{n}_{p,\uparrow} \hat{n}_{p,\downarrow}$ the interaction Hamiltonians. Then, we define the contour-ordered single particle Green's function of the system for the active and passive dots, respectively, by

$$iG^c_a(\tau, \tau') = \left\langle T_c \hat{d}_a(\tau) \hat{d}^\dagger_a(\tau') \right\rangle, \tag{S4}$$

$$i\boldsymbol{G}^c_p(\tau, \tau') = \left\langle T_c \hat{\Psi}_p(\tau) \hat{\Psi}^\dagger_p(\tau') \right\rangle, \tag{S5}$$

where $\langle \ldots \rangle$ is the ground state expectation value of the interacting system, $T_c$ is the time-ordering operator along the Keldysh contour and $\tau$ and $\tau'$ are time variables along the Keldysh contour. In Eq. (S5), we represent the Green's function of the passive dot in the Nambu basis defined by $\hat{\Psi}^\dagger_p = (\hat{d}^\dagger_{p,\uparrow}, \hat{d}_{p,\downarrow})$. Here and in the following, we show the quantities in the Nambu basis using bold letters such as $\boldsymbol{G}$ and $\boldsymbol{\Sigma}$.

## A. Dyson equation for the passive dot

In frequency space, the retarded interacting Green's function of the passive dot can be obtained from

$$\boldsymbol{G}_p^R(\omega) = \left\{ \left[\boldsymbol{g}_p^R(\omega)\right]^{-1} - \boldsymbol{\Sigma}_{p,\text{int}}^R(\omega) \right\}^{-1}, \tag{S6}$$

where $\boldsymbol{g}_p^R(\omega)$ is the mean-field retarded Green's function given by

$$\boldsymbol{g}_p^R(\omega) = \left(\omega \mathbf{I} - \boldsymbol{h}_p - \boldsymbol{\Sigma}_{p,\text{leads}}^R\right)^{-1}, \tag{S7}$$

where $\mathbf{I}$ is the $2\times 2$ identity matrix and $\boldsymbol{h}_p$ is a diagonal matrix with diagonal elements $(\varepsilon_p + U_{ap}\langle \hat{n}_a\rangle + U_p\langle \hat{n}_{p,\downarrow}\rangle, -\varepsilon_p - U_{ap}\langle \hat{n}_a\rangle - U_p\langle \hat{n}_{p,\uparrow}\rangle)$. Moreover, $\boldsymbol{\Sigma}_{p,\text{leads}}^R(\omega) = \boldsymbol{\Sigma}_{p,N}^R(\omega) + \boldsymbol{\Sigma}_{p,S}^R(\omega)$ is the sum of self-energies due to coupling the passive dot to the normal and superconducting electrodes. Their respective expressions in the wide-band approximation read

$$\boldsymbol{\Sigma}_{p,N}^R(\omega) = -i\Gamma_{p,N}\mathbf{I}, \tag{S8}$$

and

$$\boldsymbol{\Sigma}_S^R(\omega) = -i\Gamma_S \beta(\omega) \begin{pmatrix} 1 & -\frac{\Delta}{\omega} \\ -\frac{\Delta}{\omega} & 1 \end{pmatrix}, \tag{S9}$$

where $\beta(\omega)$ is given by

$$\beta(\omega) = \frac{|\omega|}{\sqrt{\omega^2 - \Delta^2}}\theta(|\omega| - \Delta) - i\frac{\omega}{\sqrt{\Delta^2 - \omega^2}}\theta(\Delta - |\omega|), \tag{S10}$$

with $\theta(\ldots)$ the Heaviside step function. Furthermore, in Eq. (S6), $\boldsymbol{\Sigma}_{p,\text{int}}^R(\omega)$ is the retarded self-energy of the passive dot due to interaction with the active dot. Its expression will be given later.

We also need to calculate the interacting lesser and greater Green's functions of the passive dot, which are given by

$$\boldsymbol{G}_p^{<,>}(\omega) = \boldsymbol{G}_p^R(\omega)\left(\boldsymbol{\Sigma}_{p,\text{leads}}^{<,>}(\omega) + \boldsymbol{\Sigma}_{p,\text{int}}^{<,>}(\omega)\right)\boldsymbol{G}_p^A(\omega), \tag{S11}$$

where the advanced Green's function is obtained as $\boldsymbol{G}_p^A(\omega) = \left[\boldsymbol{G}_p^R(\omega)\right]^\dagger$ and $\boldsymbol{\Sigma}_{p,\text{leads}}^{<,>}$ can be calculated from

$$\boldsymbol{\Sigma}_{p,N/S}^< = -2i\,\text{Im}\left(\boldsymbol{\Sigma}_{p,N/S}^R\right)f_{N/S}(\omega), \tag{S12}$$

and

$$\boldsymbol{\Sigma}_{p,N/S}^> = 2i\,\text{Im}\left(\boldsymbol{\Sigma}_{p,N/S}^R\right)[1 - f_{N/S}(\omega)]. \tag{S13}$$

$f_\beta(\omega) = \{1 + \exp[(\omega - \mu_\beta)/k_B T]\}^{-1}$ is the Fermi distribution function of electrode $\beta = N, S$ with chemical potential $\mu_\beta$ and temperature $T$. The expressions for the lesser and greater interacting self-energies, $\boldsymbol{\Sigma}_{p,\text{int}}^{<,>}$, are given below.

## B. Dyson equation for the active dot

Next, we will focus on the active dot. Its full retarded Green's function is given by

$$G_a^R(\omega) = \left\{ \left[g_a^R(\omega)\right]^{-1} - \Sigma_{a,\text{int}}^R(\omega) \right\}^{-1}, \tag{S14}$$

where the mean-field retarded Green's function $g_a^R(\omega)$ is

$$g_a^R(\omega) = \left(\omega - \varepsilon_a - U_{ap}\sum_\sigma \langle \hat{n}_{p,\sigma}\rangle - \Sigma_{a,\text{leads}}^R\right)^{-1}. \tag{S15}$$

Here, $\Sigma_{a,\text{leads}}^R(\omega) = \Sigma_{a,L}^R(\omega) + \Sigma_{a,R}^R(\omega)$ is the self-energy due to coupling the active dot to its right and left normal metal electrodes: $\Sigma_{a,\beta}^R(\omega) = -i\Gamma_{a,\beta}$ for $\beta = R, L$. The lesser Green's function of the active dot is also given in a similar manner to the passive dot in Eq. (S11) by replacing subscript $p$ with $a$ and considering the hybridization of the active dot with two normal metallic electrodes.

3## C. Interaction self-energies for the passive dot

We obtain the interacting lesser and greater self-energies for the passive dot within second-order perturbation theory [S2]. We generalize the results of Ref. [S3] to include the nonlocal capacitive interaction between the passive and active dots:

$$\boldsymbol{\Sigma}_{p,\text{int}}^{<,>}(\omega) = \left(\frac{U_p}{2\pi}\right)^2 \int d\omega_1 Q_p^{<,>}(\omega_1) \hat{\sigma}_y \left[\boldsymbol{G}_p^{>,<}(\omega_1 - \omega)\right]^T \hat{\sigma}_y$$
$$+ \left(\frac{U_{ap}}{2\pi}\right)^2 \int d\omega_1 \hat{\sigma}_z \boldsymbol{G}_p^{<,>}(\omega_1) \hat{\sigma}_z W_a^{<,>}(\omega - \omega_1), \qquad (S16)$$

where $\hat{\sigma}_y$ and $\hat{\sigma}_z$ are the second and third Pauli matrices, $Q_p$ reads

$$Q_p^{<,>}(\omega) = \int d\omega_1 \left[G_{p,11}^{<,>}(\omega_1) G_{p,22}^{<,>}(\omega - \omega_1) - G_{p,12}^{<,>}(\omega_1) G_{p,21}^{<,>}(\omega - \omega_1)\right], \qquad (S17)$$

and $W_a$ is given by

$$W_a^{<,>}(\omega) = \int d\omega_1 G_a^{<,>}(\omega_1) G_a^{>,<}(\omega_1 - \omega). \qquad (S18)$$

The interacting retarded self-energy of the passive dot is

$$\boldsymbol{\Sigma}_{p,\text{int}}^R(\omega) = \left(\frac{U_p}{2\pi}\right)^2 \int d\omega_1 \left[Q_p^<(\omega_1) \hat{\sigma}_y \left[\boldsymbol{G}_p^A(\omega_1 - \omega)\right]^T \hat{\sigma}_y + Q_p^R(\omega_1) \hat{\sigma}_y \left[\boldsymbol{G}_p^<(\omega_1 - \omega)\right]^T \hat{\sigma}_y\right]$$
$$+ \left(\frac{U_{ap}}{2\pi}\right)^2 \int d\omega_1 \left[\hat{\sigma}_z \boldsymbol{G}_p^<(\omega_1) \hat{\sigma}_z W_a^R(\omega - \omega_1) + \hat{\sigma}_z \boldsymbol{G}_p^R(\omega_1) \hat{\sigma}_z \left[W_a^<(\omega - \omega_1) + W_a^R(\omega - \omega_1)\right]\right], \qquad (S19)$$

where

$$Q_p^R(\omega) = \int d\omega_1 \left[G_{p,11}^<(\omega_1) G_{p,22}^R(\omega - \omega_1) - G_{p,12}^<(\omega_1) G_{p,21}^R(\omega - \omega_1)\right.$$
$$+ G_{p,11}^R(\omega_1) G_{p,22}^<(\omega - \omega_1) - G_{p,12}^R(\omega_1) G_{p,21}^<(\omega - \omega_1)$$
$$\left.+ G_{p,11}^R(\omega_1) G_{p,22}^R(\omega - \omega_1) - G_{p,12}^R(\omega_1) G_{p,21}^R(\omega - \omega_1)\right], \qquad (S20)$$

and

$$W_a^R(\omega) = \int d\omega_1 \left[G_a^<(\omega_1) G_a^A(\omega_1 - \omega) + G_a^R(\omega_1) G_a^<(\omega_1 - \omega)\right]. \qquad (S21)$$

## D. Interacting self-energies for the active dot

The interacting lesser and greater self-energies for the active dot are given by

$$\Sigma_{a,\text{int}}^{<,>}(\omega) = \left(\frac{U_{ap}}{2\pi}\right)^2 \int d\omega_1 G_a^{<,>}(\omega_1) W_p^{<,>}(\omega - \omega_1), \qquad (S22)$$

where the $W_p$ functions are

$$W_p^{<,>}(\omega) = \int d\omega_1 \text{Tr}\left[\hat{\sigma}_z \boldsymbol{G}_p^{<,>}(\omega_1) \hat{\sigma}_z \boldsymbol{G}_p^{>,<}(\omega_1 - \omega)\right], \qquad (S23)$$

$$W_p^R(\omega) = \int d\omega_1 \text{Tr}\left[\hat{\sigma}_z \boldsymbol{G}_p^<(\omega_1) \hat{\sigma}_z \boldsymbol{G}_p^A(\omega_1 - \omega) + \hat{\sigma}_z \boldsymbol{G}_p^R(\omega_1) \hat{\sigma}_z \boldsymbol{G}_p^<(\omega_1 - \omega)\right], \qquad (S24)$$

Tr[...] being the trace over the Nambu matrices. The interacting retarded self-energy of the active dot is given by

$$\Sigma_{a,\text{int}}^R(\omega) = \left(\frac{U_{ap}}{2\pi}\right)^2 \int d\omega_1 \left[G_a^<(\omega_1) W_p^R(\omega - \omega_1) + G_a^R(\omega_1) \left[W_p^<(\omega - \omega_1) + W_p^R(\omega - \omega_1)\right]\right]. \qquad (S25)$$

### E. Expression for the current

The above discussion provides a complete description of the required equations to calculate the interacting Green's functions of both active and passive dots in the nonequilibrium steady state. In our numerical calculations, we have performed self-consistent calculations to obtain the self-consistent Green's functions and self-energies of the system. Once the NEGFs are obtained the electric current through the active dot to the lead $\beta = L, R$ can be calculated as [S1]

$$I_{a,\beta} = \frac{e}{\hbar} \int \frac{d\omega}{2\pi} \left[ G_a^<(\omega) \Sigma_{a,\beta}^>(\omega) - G_a^>(\omega) \Sigma_{a,\beta}^<(\omega) \right]. \tag{S26}$$

On the other hand, the electric current through the passive dot to the lead $\beta = N, S$ can be evaluated from $I_{p,\beta} = \int d\omega \mathcal{I}_\beta(\omega)$, where $\mathcal{I}_\beta(\omega)$ is the energy resolved current density [S4]

$$\mathcal{I}_\beta(\omega) = \frac{e}{2h} \text{Tr} \left[ \hat{\sigma}_z (\boldsymbol{G}_p^R(\omega) \boldsymbol{\Sigma}_{p,\beta}^<(\omega) + \boldsymbol{G}_p^<(\omega) \boldsymbol{\Sigma}_{p,\beta}^A(\omega) - \boldsymbol{\Sigma}_{p,\beta}^R(\omega) \boldsymbol{G}_p^<(\omega) - \boldsymbol{\Sigma}_{p,\beta}^<(\omega) \boldsymbol{G}_p^A(\omega)) \right]. \tag{S27}$$

This expression for the electric current of the passive dot has the advantage that by taking appropriate integration limits we can quantitatively distinguish between the current in the subgap domain $(I_A)$, which is mainly due to pair tunneling accompanied by Andreev reflection, and the current outside the superconductor gap $(I_{qp})$, which is dominated by quasiparticle tunneling into the superconductor continuum. It is thus clear that the total current flowing through the passive dot can be rewritten as $I_p = I_A + I_{qp}$ where

$$I_A = \int_{-\Delta}^{\Delta} \mathcal{I}(\omega) \, d\omega, \tag{S28}$$

and

$$I_{qp} = \left( \int_{-\infty}^{-\Delta} + \int_{\Delta}^{\infty} \right) \mathcal{I}(\omega) \, d\omega. \tag{S29}$$

## S-III. RATE EQUATION METHOD

The rate equation method can be employed in the $k_B T \gg \Gamma_\beta$ regime to calculate the electric current using the eigenstates and eigenenergies of $\hat{H}_{\text{dqd}}$ as follows [S5]

$$I_\alpha = \frac{e}{\hbar} \sum_{\lambda,\kappa} \left( \gamma_{\lambda\kappa}^{\alpha N} - \Gamma_{\lambda\kappa}^{\alpha N} \right) P(\kappa), \tag{S30}$$

where $\alpha = p, a$ as before and

$$\Gamma_{\lambda\kappa}^{\alpha\beta} = \Gamma_\beta |\langle \lambda | \hat{\delta}_\alpha^\dagger | \kappa \rangle|^2 f_\beta (E_\lambda - E_\kappa), \tag{S31}$$

$$\gamma_{\lambda\kappa}^{\alpha\beta} = \Gamma_\beta \left| \langle \lambda | \hat{\delta}_\alpha | \kappa \rangle \right|^2 [1 - f_\beta (E_\kappa - E_\lambda)], \tag{S32}$$

are the tunneling rates in and out of the dot $\alpha$, from and into electrode $\beta = N, L, R, S$, respectively. In the above equations, $\hat{\delta}_a \equiv \hat{d}_a$ and $\hat{\delta}_p \equiv \sum_\sigma \hat{d}_{p,\sigma}$. Moreover, $|\lambda\rangle$ is the eigenstate of $\hat{H}_{\text{dqd}}$ with energy $E_\lambda$. In Eq. (S30), $P(\lambda)$ is the occupation probability for state $|\lambda\rangle$ which is calculated by solving the system of equations $0 = \sum_{\alpha,\beta,\kappa} \left( (\Gamma_{\kappa\lambda}^{\alpha\beta} + \gamma_{\kappa\lambda}^{\alpha\beta}) P(\lambda) - (\Gamma_{\lambda\kappa}^{\alpha\beta} + \gamma_{\lambda\kappa}^{\alpha\beta}) P(\kappa) \right)$ together with the normalization condition $\sum_\kappa P(\kappa) = 1$.

### A. Current carried by Cooper pairs

To investigate the drag current corresponding to the subgap electron transport in the passive dot, it is useful to take the infinite-gap approximation. This approach is equivalent to taking the superconductor energy gap to be the largest energy scale in the system. Hence, the coupling between the superconducting electrode and the passive dot

can be replaced with an effective pairing term $\Gamma_S(\hat{d}_{p,\uparrow}^\dagger \hat{d}_{p,\downarrow}^\dagger + \text{h.c.})$ in the Hamiltonian of the passive dot [S6, S7]. As a consequence, the Hamiltonian of the isolated double quantum dot can be exactly diagonalized as

$$\begin{aligned} |\sigma, n\rangle, & \quad E_{\sigma,n} = \varepsilon_p + n(\varepsilon_a + U_{ap}), \\ |\pm, n\rangle = \mathcal{N}_{\pm,n}^{-1}(A_{\pm,n}|0,n\rangle - \Gamma_S|2,n\rangle), & \quad E_{\pm,n} = n\varepsilon_a + A_{\mp,n}, \end{aligned} \quad (S33)$$

where the first argument in the kets (0, $\sigma$ or 2) represents the state of the passive dot while the second ($n = 0, 1$) is the occupation of the active dot. Moreover, $\mathcal{N}_{\pm,n}$ is a normalization factor and

$$A_{\pm,n} = \tilde{\varepsilon}_n \pm \sqrt{(\tilde{\varepsilon}_n)^2 + \Gamma_S^2}, \tag{S34}$$

where $\tilde{\varepsilon}_n = \varepsilon_p + U_p/2 + nU_{ap}$. The above eigensystem is composed of eight states which allows us to employ Eq. (S30) to calculate the electric current through the passive and active dots in the infinite gap approximation.

### 1. Tunneling rates and electron-hole symmetry

The number of electrons in the passive dot is not well defined when it is in one of the superposition states $|\pm, n\rangle$. Then, transitions to odd states $|\sigma, n\rangle$ may involve an electron either tunneling in or out of the dot. Their probabilities are determined by the tunneling rates $\Gamma_{\lambda\kappa}^{\alpha\beta} = \mathcal{G}_{\lambda\kappa}^{\alpha\beta} f_\beta(E_\lambda - E_\kappa)$ for transitions involving an electron tunneling from terminal $\beta$ into quantum dot $\alpha$, and $\gamma_{\lambda\kappa}^{\alpha\beta} = \mathcal{J}_{\lambda\kappa}^{\alpha\beta}[1 - f_\beta(E_\kappa - E_\lambda)]$ for those involving an electron tunneling out of the dot. Here, $\mathcal{G}_{\lambda\kappa}^{\alpha\beta} = \Gamma_\beta |\langle\lambda|\hat{\delta}_\alpha^\dagger|\kappa\rangle|^2$ and $\mathcal{J}_{\lambda\kappa}^{\alpha\beta} = \Gamma_\beta |\langle\lambda|\hat{\delta}_\alpha|\kappa\rangle|^2$.

For the transition $|\sigma, n\rangle \to |+, n\rangle$, we have

$$\mathcal{G}_{+n,\sigma n}^{p,N} = \Gamma_N \mathcal{N}_{+n}^{-2} \Gamma_S^2 \tag{S35}$$

$$\mathcal{J}_{+n,\sigma n}^{p,N} = \Gamma_N \mathcal{N}_{+n}^{-2} \left(\tilde{\varepsilon}_n + \sqrt{\tilde{\varepsilon}_n^2 + \Gamma_S^2}\right)^2. \tag{S36}$$

Whether this transition is more likely to happen with an electron tunneling in or out of the passive dot depends on the sign of $\tilde{\varepsilon}_n$. The two rates are equal at $\tilde{\varepsilon}_n = 0$.

For $|-, n\rangle \to |\sigma, n\rangle$, we have

$$\mathcal{G}_{\sigma n,-n}^{p,N} = \Gamma_N \mathcal{N}_{-n}^{-2} \left(-\tilde{\varepsilon}_n + \sqrt{\tilde{\varepsilon}_n^2 + \Gamma_S^2}\right)^2 \tag{S37}$$

$$\mathcal{J}_{\sigma n,-n}^{p,N} = \Gamma_N \mathcal{N}_{-n}^{-2} \Gamma_S^2. \tag{S38}$$

Note that at the point $\tilde{\varepsilon}_0 + \tilde{\varepsilon}_1 = 0$, the electron-hole symmetry is established by having $\mathcal{G}_{\pm 1,\sigma 1}^{p,N} = \mathcal{J}_{\sigma 0, \mp 0}^{p,N}$. Hence, at this point, sequences of the form $|\sigma, 0\rangle \to |+, 0\rangle \to |-, 1\rangle \to |\sigma, 1\rangle$ contribute on average to the transport of an electron and of a hole. The same (though with opposite contributions) is valid for the cycle $|\sigma, 1\rangle \to |+, 1\rangle \to |-, 0\rangle \to |\sigma, 0\rangle$, resulting in no drag current.

### 2. NEGF vs. rate equation results

Our NEGF formalism is capable of considering the coupling between the passive dot and the leads nonperturbatively. However, it is expected that in the weak tunneling regime, where $k_B T \gg \Gamma_\beta$, the NEGF results reproduce the results of the rate equation method.

The infinite-gap approximation which we discussed earlier in this section can be explored within the NEGF formalism by setting a large value to $\Delta$ in Eq. (S9). In Fig. S1, we compare the results obtained from both methods. In Fig. S1(a), we consider a large bias voltage on the active dot and plot the drag current as a function of $\varepsilon_p$. We can see that the NEGF results are in good agreement with those of the rate equation method, especially the sign of the drag current as a function of $\varepsilon_p$, and the inversion of the drag current at the point $\varepsilon_p = -(U_{ap} + U_p)/2 \equiv -U_{\text{total}}/2$. In Fig. S1(b), we take $\varepsilon_p = -0.7 U_{\text{total}}$, and plot the drag current as a function of bias voltage on the active dot, where we find good agreement between the results obtained from both methods. The insets in both panels show the corresponding results for the case where $k_B T \sim \Gamma_N$. We can see that by decreasing the temperature the rate equation results depart from the NEGF ones, which is an indication that in this parameter regime the results from rate equation method are not quantitatively correct, though they still give a proper qualitative behaviour.

6

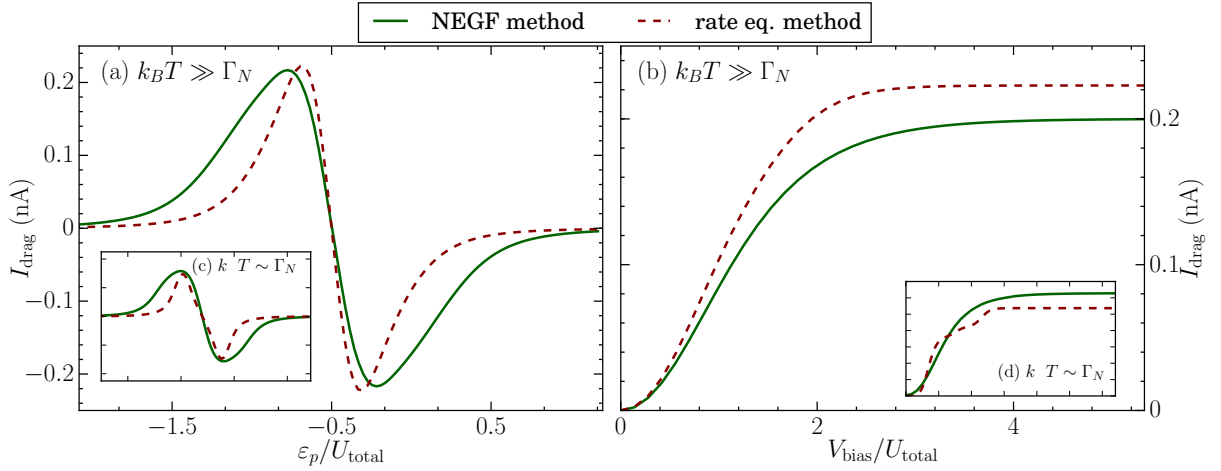

FIG. S1. Comparison of NEGF results with the results obtained from rate equation method in the infinite gap approximation. Left panels show drag current as a function of $\varepsilon_p$ for a large $V_{\text{bias}}$ while right panels show drag current as a function of $V_{\text{bias}}$ for $\varepsilon_p = -0.7 U_{\text{total}}$. In (a) and (b) we take $k_B T = 1$ meV whereas in (c) and (d) $k_B T = 0.03$ meV. Here, $\Delta$ in the NEGF calculations is set to a large value ($\Delta = 10$ meV). Additional parameters: $U_p = 5 U_{ap} = 0.5$ meV and $\Gamma_L = \Gamma_R = \Gamma_N = \Gamma_S/3 = 0.05$ meV.

### B. Current carried by quasiparticles

Let us now consider the case when the contribution of Cooper pairs is negligible. The only effect of the superconductor is then introduced by the presence of a gap in the density of states. Tunneling through the $S$ barrier is only possible for electrons with energy falling outside the gap region.

Assuming strong on-site Coulomb interactions, there are only four relevant charge states, described in terms of the charge occupations: $|n_p, n_a\rangle$, with $n_\alpha = 0, 1$. We ignore the spin degree of freedom here, for simplicity. The different tunneling rates in the passive dot are:

$$\Gamma^{p,N}_{1n,0n} = \Gamma_N f_{Nn}, \tag{S39}$$

for electrons tunneling in from terminal $N$ with the active dot having $n$ electrons, and

$$\Gamma^{p,S}_{1n,0n} = \Gamma^{\text{qp}}_{Sn} f_{Sn} = \Gamma_S \nu_n f_{Sn}, \tag{S40}$$

for electrons tunneling in from terminal $S$. For transitions into the active dot, we have $\Gamma^{a,\beta}_{n1,n0} = \Gamma_\beta f_{\beta n}$. Here, $f_{\beta n} = f_\beta(\varepsilon_p + n U_{ap})$, when $\beta = N, S$, and $f_{\beta n} = f_\beta(\varepsilon_a + n U_{ap})$, when $\beta = L, R$. For electrons tunneling out to the respective terminals, we need to make the replacement $f_{\beta n} \to 1 - f_{\beta n}$. Since the passive system is unbiased, we define $f_n \equiv f_{Nn} = f_{Sn}$. This way, tunneling events involving the superconductor depend on the occupation of the active dot. Note that we are assuming the simplest case where $\Gamma_N$ and $\Gamma_S$ are energy-independent, which emphasizes the key role of the gap.

Writing down a rate equation for these states [S8], we find the drag current

$$I_{\text{drag}} = -e(\nu_0 - \nu_1) \frac{\Gamma_N \Gamma_S}{\hbar \gamma^3} \sum_{\beta, \beta' = L, R} \mathcal{A}_{\beta \beta'} \Gamma_\beta \Gamma_{\beta'}, \tag{S41}$$

with the prefactor $\gamma^3 > 0$ setting the normalization of the steady-state density matrix, and

$$\mathcal{A}_{\beta \beta'} = f_0 f_{\beta 1}(1 - f_{\beta 0}) - f_1 f_{\beta 0}(1 - f_{\beta' 1}) + \frac{1}{2}(f_{\beta 0} + f_{\beta' 0} - f_{\beta 1} - f_{\beta' 1}). \tag{S42}$$

Remarkably, the presence of a drag current only relies on the asymmetry introduced by the gap, $\nu_0 - \nu_1$.

We obtain a simpler expression in the high bias limit, where $f_{Ln} \to 1$ and $f_{Rn} \to 0$, hence tunneling is unidirectional in the active system and:

$$I_{\text{drag}} = -e \frac{(\nu_0 - \nu_1)(f_0 - f_1)\Gamma_N \Gamma_S \Gamma_L \Gamma_R}{(\Gamma_L + \Gamma_R)\{\Gamma_N[\Gamma_N + (\nu_0 + \nu_1)\Gamma_S + \Gamma_L + \Gamma_R] + \Gamma_S(\nu_0 \nu_1 \Gamma_S + \nu_1 \Gamma_L + \nu_0 \Gamma_R)\}}. \tag{S43}$$

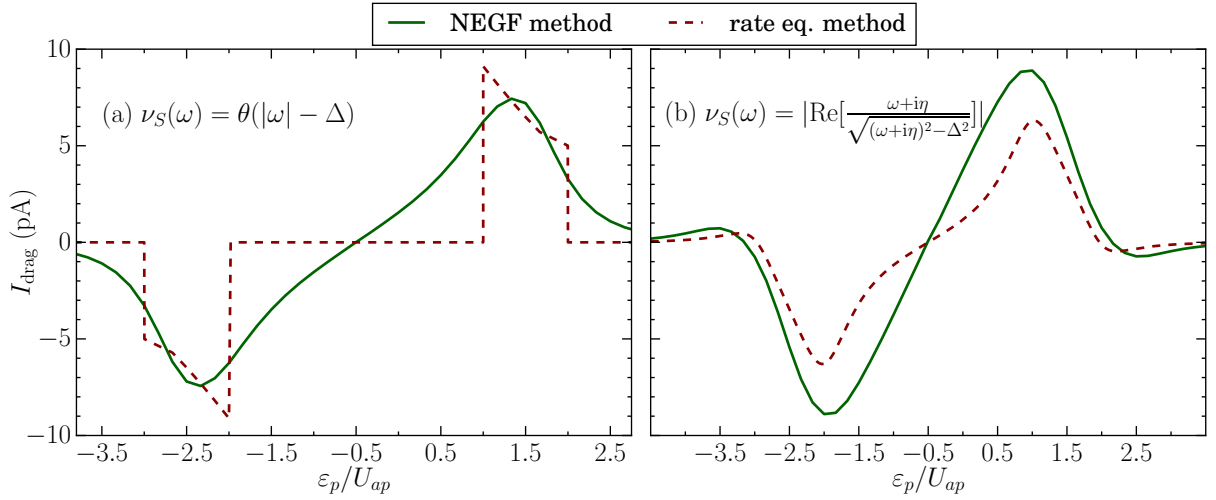

FIG. S2. Drag current as a function of $\varepsilon_p$ calculated using the NEGF (solid line) and the rate equation (dashed line) methods, when the superconductor lead is replaced by a normal metal with a modified density of states, $\nu_S(\omega)$. We take in (a) a flat pseudo-gap and in (b) a Dynes density of states with $\eta = \Delta/6$ for the normal lead. Additional parameters: $\Delta = 1$ meV, $\Gamma_L = \Gamma_R = \Gamma_N = \Gamma_S = 0.0016$ meV, $U_p = 0$ and $U_{ap} = k_B T = 0.5$ meV.

We find a simple interpretation from the above expression for the drag current by introducing a hard gap in the density of states $\nu_n = g(\varepsilon_p + nU_{ap})\theta(|\varepsilon_p + nU_{ap}| - \Delta)$. For the moment, we ignore the explicit energy dependence of the function $g(E)$. Now, consider for example that the passive dot level lies in the gap of the superconductor, $-\Delta < \varepsilon_p < \Delta$, but $\varepsilon_p + U_{ap} > \Delta$. Then, $\nu_0 = 0$ and $\nu_1 = 1$. Further, consider for simplicity that $k_B T \ll U_{ap}$, thereby we can approximate $f_0 \to 1$ and $f_1 \to 0$. Then, the only possible way to charge the passive dot is by an electron tunneling from $N$ when the active dot is empty. Due to the gap, this electron cannot tunnel out until the active dot becomes charged, in which case it can tunnel over the gap into the superconductor. This sequence is completed when the active dot returns to its empty state, hence generating the drag effect. Note that the electron in the passive dot might also tunnel back to $N$, in this case not contributing to the current. For finite temperatures, this sequence is still the dominant process as $f_0 - f_1 > 0$. Therefore, the drag current is positive.

The level position with respect to the gap changes the sign of the current. In the opposite case when $-\Delta < \varepsilon_p + U_{ap} < \Delta$ and $\varepsilon_p < -\Delta$ (i.e., $\nu_0 \neq 0$ and $\nu_1 = 0$), the passive dot can be charged both from $N$ and $S$, but it can only tunnel out to $N$, hence leading to a negative drag current. If both energies lie in the gap, then $\nu_0 = \nu_1 = 0$ and there is no drag current.

Thus, for levels close to the Fermi energy the direction of the current is strongly dominated by the gap. Only in the case where both energies fall outside the gap will the sign of the current depend on details of the density of states through the function $g(E)$.

### 1. NEGF vs. rate equation results

We can also compare the quasiparticle drag current obtained from the rate equation formalism with the NEGF results. To this end, we replace the superconductor lead self-energy in Eq. (S9) with a normal lead self-energy as in Eq. (S8) multiplied by an energy dependent density of states, $\nu_S(\omega)$.

In Fig. S2(a), we show the results obtained from both NEGF and rate equation methods for a hard gap density of states $\nu_S(\omega) = \Theta(|\omega| - \Delta)$. In this case, the rate equation expression calculated from Eq. (S43) is able to predict the correct sign of the drag current as a function of $\varepsilon_p$, as compared to the NEGF result. However, the rate equation results show features with sharp edges, which differ from the smooth NEGF curves. We can understand this because the rate equation method neglects quantum dot level broadenings due to tunneling. Hence, for piecewise constant pseudo-gap conditions, the drag current is conditioned on $\varepsilon_p$ and $\varepsilon_p + U_{ap}$ laying on different parts of the $\nu(E)$ profile (one in and one out of the gap).

In order to correct this and introduce finite quasiparticle lifetimes, one typically considers the Dynes density of

states [S9, S10] which is given by

$$\nu(\omega) = \left| \text{Re} \left[ \frac{\omega + i\eta}{\sqrt{(\omega + i\eta)^2 - \Delta^2}} \right] \right|, \tag{S44}$$

where $\eta$ is a positive constant. Figure S2(b) shows the drag current obtained from both methods using Eq. (S44). We observe that the sign of the drag current is again correctly reproduced by both methods while the broadening of the drag current is much better in this case. We remark the additional drag current sign change outside the gap due to the energy dependence of the Dynes density of states near the gap edges.

## ACKNOWLEDGMENTS



\end{document}